\documentclass[acmsmall, screen, nonacm]{acmart}
\usepackage{proof}
\usepackage{mathpartir}
\usepackage[frozencache,cachedir=.]{minted}
\usepackage{nccmath}

\newcommand{\rcd}[1]{\{#1\}}
\newcommand{\tint}{\text{int}}
\newcommand{\tabs}{\text{Abs}}
\newcommand{\tpre}{\text{Pre}}
\newcommand{\with}{\text{ with }}
\newcommand{\rcdX}{\rcd{x = 1}}
\newcommand{\rcdY}{\rcd{y = 2}}
\newcommand{\withY}{r \with \rcdY}
\newcommand{\addY}{\lambda r.\ \withY}
\newcommand{\seltype}{\rcd{y : \tint, x : \tint}}
\newcommand{\absLeft}{\rcd{y : \tabs, x : \tint}}
\newcommand{\absType}{\absLeft \rightarrow \seltype}
\newcommand{\rcdXType}{\rcd{x : \tint}}

\newcommand{\extT}{\rcd{y : \tint, \rho}}
\newcommand{\absT}{\alpha \rightarrow \extT}
\newcommand{\appfT}{\rcdXType \rightarrow \beta}
\newcommand{\expT}{\rcd{ y : \theta, x : \gamma, \rho'}}
\newcommand{\expandedT}{\rcd{y : \tint, x : \gamma, \rho'}}
\newcommand{\selRcdT}{\rcd{x : \delta, \rho'{}'}}
\newcommand{\expandedSelRcdT}{\rcd{x : \delta, y : \epsilon, \rho'{}'{}'}}
\newcommand{\pluseq}{\mathrel{{+}{=}}}
\newcommand{\tab}{\hspace{2pt}\vrule}

\newcommand{\typing}[1][ ]{\Gamma #1 \vdash}
\renewcommand{\infer}[1][ ]{\Sigma {#1} \vdash}

\begin{document}

\title{Towards Algebraic Subtyping for Extensible Records}

\author{Rodrigo Marques}
\author{Mário Florido}
\author{Pedro Vasconcelos}

\affiliation{%
  \institution{LIACC, Departamento de Ciência de Computadores, Faculdade de Ciências, Universidade do Porto}
  \streetaddress{rua Campo Alegre s/n}
  \postcode{4169-007}
  \city{Porto}
  \country{Portugal}
}

\renewcommand{\shortauthors}{R. Marques, P. Vasconcelos and M. Florido}

\begin{abstract}
MLsub is a minimal language with a type system combining subtyping and parametric polymorphism and a type inference algorithm which infers compact principal types. Simple-sub is an alternative inference algorithm which can be implemented efficiently and is easier to understand. MLsub supports explicitly typed records which are not extensible. Here we extend Simple-sub with extensible records, meaning that we can add new fields to a previously defined record. For this we add row variables to our type language and extend the type constraint solving method of our type inference algorithm accordingly, keeping the decidability of type inference.
\end{abstract}


\maketitle

\begin{acks}
This work is supported by the~\emph{Funda\c{c}\~ao para a Ci\^encia e Tecnologia, I.P.} (\url{https://www.fct.pt})
under grant numbers~{UIDB/00027/2020} and~{UIDP/00027/2020}.
\end{acks}

\section{Introduction}

MLsub \cite{Mlsub} is a minimal ML-like functional language with records that successfully combines algebraic subtyping and parametric polymorphism \cite{DamasMilner}. It has decidable type inference and the type inference algorithm is sound and complete, in the sense that it returns principal types. Simple-sub \cite{SimpleSub} is a subsequent formulation of a type inference algorithm which has the same features but is easier to understand because it does not require any concepts from abstract algebra. 

Inspired by previous work on extensible records by Rémy and Pottier \cite{Remy93, Pottier98}, we extend Simple-sub with record extension, adding row variables to record types. We present a new definition of subtyping, a new type system and a new type inference algorithm to deal with these additions.

A prototype implementation and code of our type inference algorithm can be found in the Github repository:
\urlstyle{tt}
\url{https://github.com/RodrigoMarques16/simple-sub-records}.

This is ongoing work. We wish to apply our inference algorithm to more elaborate examples and pursue formal proofs of soundness and completeness that we believe will hold, but do not have yet done. However, we believe that our extension, particularly the new notion of subtyping is an important step towards the integration of row variables in a new type system with extensible records and algebraic subtyping, keeping all the known advantages of Simple-sub.

\section{Terms and Types}

We now briefly review the term and type languages used by Simple-sub \cite{SimpleSub}. We will only present the syntax, type rules and subtyping cases which are going to be used in our extension for extensible records.
Terms include integers, variables, lambda abstraction, application, records and let expressions. The type language consists of base types, type variables, functional types, records, top $\top$ (the type all values), bottom $\bot$ (the type of no values), union and intersection types, and recursive types. 

\begin{align*}
    t &::= n\
    |\ x\ 
    |\ \lambda x. t\ 
    |\ t\ t\ 
    |\ \{ l_0 = t\ ,\ ...\ ,\ l_n = t \}\ 
    |\ t.l\
    |\ \text{let rec } x = t  \text{ in } t
    & \text{terms}
    \\
    \tau &::= \tint\
    |\ \tau \rightarrow \tau\
    |\ \{ l_0: \tau\ ,\ ...\ ,\ l_n: \tau \}\
    |\ \alpha\
    |\ \top |\ \bot\
    |\ \tau \sqcup \tau\
    |\ \tau \sqcap \tau\
    |\ \mu \alpha . \tau
    & \text{types}
\end{align*}


The type system of Simple-sub follows; for brevity we present only a selected set of rules: typing of records, projection and the subtyping rule.

\begin{gather*}
    \inferrule[T-Rcd]
        {\typing \overline{t_i : \tau_i}^i}
        {\typing \rcd{\overline{l_i = t_i}^i} : \rcd{\overline{l_i : \tau_i}^i}}
    \qquad\qquad
    \inferrule[T-Proj]
        {\typing t : \rcd{l : \tau} }
        {\typing t.l : \tau}
    \qquad\qquad
    \inferrule[T-Sub]
        {\typing t : \tau_1 \\ \tau_1 \leq \tau_2}
        {\typing t : \tau_2}
\end{gather*}

Finally, we recall the equivalence of record types and the intersection of their corresponding field types, and also the subtyping rules for records and intersections:

\begin{gather*}
    \inferrule[S-And]
        {\forall i,\exists j, \infer \tau_j \leq \tau'_i}
        {\infer \sqcap_j\tau_j \leq \sqcap_i\tau'_i}
    \qquad
    \inferrule[S-Rcd]
        { }
        {\rcd{\overline{l_i: \tau_i}^i} \equiv \sqcap_i \rcd{l_i: \tau_i}}
    \qquad
    \inferrule[S-Depth]
        {\infer \tau_1 \leq \tau_2}
        {\infer \rcd{l:\tau_1} \leq \rcd{l:\tau_2}}
\end{gather*}




\section{Extensible Records}

Wand \cite{Wand87} introduced row polymorphism to handle inference in a system with record extension. Rémy iterates on this in \cite{Remy93}. Rémy and Pottier study subtyping and row polymorphism in \cite{Remy95, Pottier98}.
Here we present an extension to Simple-sub, combining row variables with subtyping.
First we extend the term language with an action to extend records and add rows and field types to the type language.

\begin{align*}
    t\ &::= \cdots\ |\ t \with \rcd{l = t} & \text{terms} 
    \\
    \tau\ &::= \cdots\ |\ \rcd{\rho} & \text{types} 
    \\
    \rho\ &::=\ l:\theta,\ \rho\ |\ \tabs\ |\ \alpha & \text{rows}
    \\
    \theta\  &::=\ \alpha\ |\ \tabs\ |\ \ \tpre~\tau & \text{fields}
\end{align*}

Records are built incrementally with rows, starting from the empty row {\em Abs}. We still interpret records as a sequence of fields, but one that can now be open, ending in a row variable (a placeholder for fields that are not explicitly named but could be present), or closed and ending in {\em Abs} (all implicit fields are absent). 
A field type is {\em Pre $\tau$} if it is explicitly present. {\em Abs} (absent) means that a field is not present in the record. A type variable stands for either option. 
We now extend the type system in \cite{SimpleSub} to support these new constructs. 

\begin{gather*}
    \inferrule[T-Ext]
        {\typing t_1 : \{ l : \alpha ;\ \rho \} 
            \\ \Gamma \vdash t_2 : \tau 
            } 
        {\typing t_1 \with \{l = t_2\}: \{ l : Pre\ \tau ; \rho \}}
    \qquad
    \inferrule[T-Proj]
        {\typing t : \{l : Pre\ \tau ; \rho \}}
        {\typing t.l : \tau}
\end{gather*}

Type equivalence is defined modulo commutativity of fields and {\em Abs} unfolding. Thus
we add the following axioms to the subtyping relation: 

\begin{equation*}
    \rcd{ ... ; Abs} \equiv \rcd{ ..., l : Abs ; Abs}
    \qquad
    \rcd{ l_1 : \tau_1 , l_2 : \tau_2 ; \rho} \equiv \rcd{ l_2 : \tau_2 , l_1 : \tau_1 ; \rho}
\end{equation*}

Finally we update the definition of subtyping, changing only the S-Rcd and S-Depth rules to account for row variables and introducing two new rules to handle field types.

\begin{gather*}
    \inferrule[S-Ignore]
        { }
        { Pre\ \tau \leq Abs}
    \qquad
    \inferrule[S-Pre]
        {\infer \tau_1 \leq \tau_2}
        {\infer Pre\ \tau_1 \leq  Pre\ \tau_2}
    \\
    \inferrule[S-Rcd]
        { }
        { \rcd{\overline{l_i : \tau_i}^i, Abs} \equiv \sqcap_i \rcd{l_i : \tau_i, Abs} }
    \qquad
    \inferrule[S-Depth]
        {\infer \tau_1 \leq \tau_2}
        {\infer \{ l : \tau_1, Abs\} \leq \{l : \tau_2, Abs\} }
\end{gather*}

The last concept we need for type inference is row {\em expansion} \cite{Pottier98}. Consider the following constraint:

\begin{gather*}
    \rcd{x : \tau, \rho} \leq \rcd{\rho'}
\end{gather*}

Constraint solving creates an expansion $\rho' \longrightarrow x : \alpha, \rho''$ and constraints $\tau \leq \alpha$ and $\rho \leq \rho''$, thus,  later occurrences of $\rho'$ are now related to label $x$ and type $\tau$ as a lower bound of its possible field types.

\section{Examples}

Consider the extension function f $=\lambda r.\ r \with \rcd{y = 1}$.
Using subtyping alone function $f$ admits a type $\rcd{} \rightarrow \rcd{ y : \tint}$
which loses all information about any existing fields of $r$.
If we then tried to infer a type for f $\ (\rcd{x=1}).x$ we would fail since the information about the presence of $x$ was lost.
With row variables we infer $\rcd{y : \beta, \rho} \rightarrow \rcd{ y : \tint, \rho}$.
The presence of the field $x$ is recorded in the row variable $\rho$ during inference so we can type this application.
We now present a type derivation tree and show how our type inference algorithm works for this example. 
Due to space limitations, we will omit {\em Pre} from fields and {\em Abs} from rows. The type derivation tree is:

\begin{equation*}
    \inferrule
        {\inferrule
            {\inferrule
                {\inferrule
                    {\inferrule{}{\typing r : \absLeft \\ 2 : \tint}}
                    {\inferrule{}{\typing[, r : \absLeft] \withY : \seltype}}}
                {\typing \addY : \absType}
                \\ 
                {\inferrule
                    {\inferrule
                        {1 : \tint}
                        {\typing \rcdX : \rcdXType}}
                    {\typing \rcdX : \absLeft}}}
            {\typing (\addY) \rcdX : \seltype}}
        {\typing ((\addY) \rcdX) . x : \tint}
\end{equation*}

Running our type inference algorithm for this term produces a trace as shown in figure \ref{inference-listing}. Function $constrain$ calls our type constraint solver. Constraints on type variables update their bounds in the environment. We omit a few steps unnecessary to infer the type of the term.

\begin{figure}
\noindent\begin{minipage}[b]{.45\textwidth}
\begin{minted}[mathescape,escapeinside=||,linenos]{c}
typeTerm |$((\addY) \rcdX) . x$|
|\tab|  typeTerm |$(\addY) \rcdX$|
|\tab|  |\tab|  typeTerm |$\addY$|
|\tab|  |\tab|  |\tab|  typeTerm |$\withY$|
|\tab|  |\tab|  |\tab|  |\tab|  constrain |$\alpha \leq \rcd{y : \theta, \rho}$|
|\tab|  |\tab|  |\tab|  = |$\extT$|
|\tab|  |\tab|  = |$\absT$|
|\tab|  |\tab|  typeTerm |$\rcdX$| = |$\rcdXType$|
|\tab|  |\tab|  constrain |$\absT \leq \appfT$|
|\tab|  |\tab|  |\tab|  constrain |$\rcdXType \leq \alpha$|
|\tab|  |\tab|  |\tab|  |\tab|  constrain |$\rcdXType \leq \rcd{y : \theta, \rho}$|
|\tab|  |\tab|  |\tab|  |\tab|  |\tab|  constrain |$\rcdXType \leq \expT$|
|\tab|  |\tab|  |\tab|  |\tab|  |\tab|  |\tab|  constrain |$\tint \leq \gamma$|
|\tab|  |\tab|  |\tab|  |\tab|  |\tab|  |\tab|  constrain |$\tabs \leq \theta$|
|\tab|  |\tab|  |\tab|  constrain |$\extT \leq \beta$|
|\tab|  |\tab|  |\tab|  |\tab|  constrain |$\expandedT \leq \beta$|
|\tab|  = |$\beta$|
|\tab|  constrain |$\beta \leq \selRcdT$|
|\tab|  |\tab|  constrain |$\expandedT \leq \selRcdT$|
|\tab|  |\tab|  |\tab|  constrain |$\expandedT \leq \expandedSelRcdT$|
|\tab|  |\tab|  |\tab|  |\tab|  constrain |$\gamma \leq \delta$|
|\tab|  |\tab|  |\tab|  |\tab|  |\tab|  constrain |$\tint \leq \delta$|
= |$\delta$|
Bounds = |$\delta \leq \tint$|
\end{minted}
\end{minipage}
\begin{minipage}[b]{.45\textwidth}
\begin{minted}[mathescape,escapeinside=||]{c}
|\newline|

            // Env $\pluseq r : \alpha$, $\alpha$ fresh

            // $\rho$, $\theta$ fresh



            // $\beta$ fresh

            // Constrain $\alpha$'s upperbounds
            // x missing, expand $\rho$


            // $\rho \longrightarrow \expandedT$



            // Constrain $\beta$'s lowerbounds
            // $y$ missing, expand $\rho''$

            // Constrain $\gamma$'s lowerbounds

|\newline|
\end{minted}
\end{minipage}
\caption{Tracing of the type inference example}
\label{inference-listing}
\end{figure}

After coalescence~\cite{SimpleSub} the type is {\em int}. Type inference for the application constrains the left term to be a function that receives and produces a $\rcd{\rho}$ (lines 9-16). The expansion of $\rho$ in line 12 is how we remember the type of $x$ later in lines 19-21.

\bibliographystyle{ACM-Reference-Format}
\bibliography{references}

\end{document}